\documentclass[showkeys]{revtex4}

\usepackage[T1]{fontenc}
\usepackage[latin1]{inputenc}

\usepackage{graphicx}
\usepackage{amsmath}

\newcommand{\ket}[1]{|#1 \rangle }
\newcommand{\bra}[1]{\langle #1 |}

\begin{document}

\title{Is Fresnel Optics Quantum Mechanics in Phase Space?}

\author{O.~Crasser, H.~Mack and W.~P.~Schleich}
\affiliation{Abteilung f\"ur Quantenphysik, Universit\"at Ulm, 89069 Ulm, Germany}


\begin{abstract}
We formulate and argue in favor of the following conjecture: There exists an intimate connection between Wigner's quantum mechanical phase
space distribution function and classical Fresnel optics. 
\end{abstract}

\maketitle

\section{Introduction}

There exists a well-established and close analogy between classical
optics and wave mechanics. Indeed, the paraxial approximation of the
Helmholtz wave equation leads \cite{key-1} to a Schr\"{o}dinger
equation. In the present article we argue that there might be another
intimate relation between quantum mechanics and wave optics. The agents
of this proposed connection are Wigner's phase space distribution function 
\cite{wigner} and Fresnel optics \cite{r3}.

The Wigner function \cite{r10} allows a formulation of quantum mechanics, which
displays a close resemblance to classical statistical mechanics. It
also provides the quantum mechanical probability distributions of
variables such as position and momentum as marginals. Moreover, the
Radon transform \cite{r10} expresses the Wigner function in terms of measurable
rotated quadrature distributions, that is of a continuous set of tomographic cuts through
the Wigner function. It is in this context
that we recall that the wavefunction in rotated quadrature variables
is related \cite{r12} to the wavefunction in the original variable
by a Fresnel transform. This fact is the first hint for a connection between the
Wigner function and Fresnel optics \cite{r13}.

However, we suspect that there must be even closer ties between these
topics. We present arguments supporting this conjecture but have to
admit that we still lack a rigorous proof.

Nevertheless, we offer these ideas as a tribute to our long term friend
Frank Moss on the occasion of his 70\textsuperscript{th} birthday. From the creation
of quantum vortices by a moving object in superfluid \(^{4}\textrm{He}\)
\cite{r16} via noise in complex systems \cite{r4,r5} to neuro dynamics
\cite{r17} Frank has covered many topics in physics. He has made
many path breaking discoveries and has always been open to new ideas.
It is only for this reason that we have the courage to propose an
idea that has not been developed to its fullest extent.

Our paper is organized as follows. In section \ref{sec2} we briefly
summarize the essential features of Wigner's phase space distribution.
Here we emphasize a crucial property that expresses the value of the
Wigner function at a given point in phase space as an alternating
sum of the energy statistics. An analogous relation exists in Fresnel
optics. In section \ref{sec3} we review a representation of the field
in the Fresnel approximation as an alternating sum of inclination
factors. We dedicate sections \ref{sec4} and \ref{sec5} to compare
and contrast the two phenomena and elaborate on a possible connection
between them. Section 6 summarizes our conclusions.

\section{Wigner Function: Alternating Sum of Occupation Probabilities\label{sec2}}

In classical mechanics the state of a system and its future is completely determined when we specify its initial position \emph{and} momentum. Hence, a single point in phase space suffices to characterize a classical system.

In contrast, the uncertainty principle prevents such a description
of a quantum system. According to Max Planck a quantum state takes
up at least an area of \(2\pi \hbar \) and the internal structure
of the state follows from a quantum mechanical phase space distribution.

There is an infinite number \cite{r10} of such distributions. This
feature is closely related to the question of operator ordering in
quantum mechanics. All quantum mechanical distribution functions contain
the same amount of information but bring out different features of
quantum mechanics.

A rather popular phase space function is the one \cite{wigner} proposed by Eugene
Paul Wigner in 1932 in order to understand quantum corrections to
a thermodynamic equilibrium. It is interesting to note that it already appeared
earlier in papers by Paul Adrian Maurice
Dirac \cite{dirac} and Werner Heisenberg \cite{heisenberg} on the Thomas-Fermi atom and on X-ray scattering from crystals,
respectively.

The recent years have seen a renaissance of the Wigner function. In
particular, the field of quantum optics has profited immensely from
the Wigner function since it brings out most clearly the superposition
principle in quantum mechanics. Moreover, Wigner functions of quantum
states have been reconstructed \cite{r14} for various quantum systems,
such as a single mode of the electromagnetic field in a cavity, an
ion stored in a Paul trap, and a Rydberg electron.

The Wigner function \(W(x,p)\) of a density operator \(\hat{\rho }\)
follows from the Fourier integral \cite{r10}\begin{equation}
\label{wigner_def}
W(x,p) =\frac{1}{2\pi \hbar }\!\int dy\, e^{-ipy/\hbar }\bra {x+{\textstyle \frac{1}{2}}y}\hat{\rho }\ket {x-{\textstyle \frac{1}{2}}y}
\end{equation}
of the matrix elements \(\rho(x',x'')\equiv\bra{x'}\hat{\rho }\ket {x''}\)
of the density operator in position representation and \(p\) denotes the conjugate variable of position \(x\).

For the case of a harmonic oscillator the Wigner function at a phase space point
\(\alpha(x,p) \equiv \kappa x+ip/\left( \hbar \kappa \right) \), where
\(\kappa \) is the characteristic length scale of the oscillator,
can be expressed as the alternating sum \cite{r9}\begin{equation}
\label{w_alpha}
W(\alpha )\equiv 2\sum ^{\infty }_{n=0}(-1)^{n}P_{n}(-\alpha ).
\end{equation}
In this formula the energy distribution \begin{equation}
\label{pn}
P_{n}(\alpha )\equiv \bra {n}\hat{D}(\alpha )\hat{\rho }\hat{D}^{\dagger }(\alpha )\ket {n}
\end{equation}
 of the displaced quantum state \(\hat{\rho }\) enters.
Here \(\ket {n}\) and \(\hat{D}(\alpha )\) denote the energy
eigenstates and the displacement operator, respectively.

We conclude by emphasizing that the relation Eq.~(\ref{w_alpha}) holds true even
for arbitrary but symmetric binding potentials.

\section{Fresnel Optics: Alternating Sum of Inclination Factors\label{sec3}}

\begin{figure}
\centering{\includegraphics{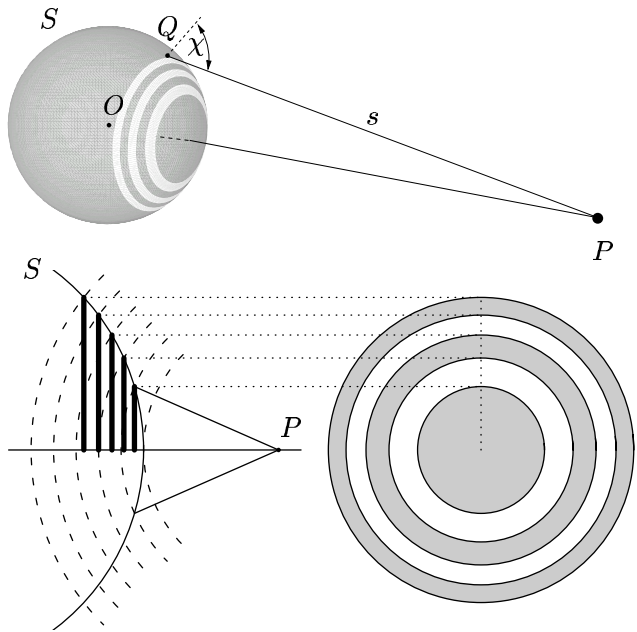}}
\caption{\label{fig1}Huygens-Fresnel principle analyzed with the help of Fresnel zones. Spheres, whose radii differ by half a wavelength and whose centers are all located at $P$, cut a spherical wavefront $S$ originating from a point $O$. The radii of the Fresnel zones indicated by solid lines in the lower part follow approximately a square root law.}
\end{figure}

The representation Eq.~(\ref{w_alpha}) of the Wigner function as
an alternating sum has served as an extremely useful tool to reconstruct
\cite{r2} this phase space function. It is also the starting point
of our expedition to Fresnel optics.

In the top of Fig.~\ref{fig1} we show a spherical wavefront of radius \(r_{0}\)
and wavenumber \(k\) that has originated from a source at \(O\). At a unit distance from
the source its amplitude is \(A\). What is the strength of this
wave at a given point \(P\) appropriately away from the source?

The Huygens-Fresnel principle expresses the strength \(\mathcal{U}(P)\)
of the wave as the integral \cite{r3}
\begin{equation}
\label{u_def}
\mathcal{U}(P)=A\frac{e^{ikr_{0}}}{r_{0}}\int_{S} \frac{e^{iks}}{s}\mathcal{K}(\chi )\, dS
\end{equation}
where $dS$ is the surface element at the point $Q$ on the spherical
wavefront and \(s\) is the distance between $P$ and $Q$. The integration extends over the surface of the spherical wave. Moreover, \(\mathcal{K}(\chi )\) is an inclination
factor and \(\chi \) is the diffraction angle, that is, the angle
between the normal at \(Q\) and the direction of the connection
between \(Q\) and \(P\).

At this stage of the argument we cannot recognize any relation between
the Wigner phase space function, Eq.~(\ref{wigner_def}), and Fresnel
optics, Eq.~(\ref{u_def}). However, the geometrical representation
of the Huygens-Fresnel integral, that is, the construction \cite{r3}
of the familiar Fresnel zones provides the crucial clue.

In Fig.~\ref{fig1} we start from a sphere around \(P\) with a radius equal to the distance between \(P\) and the point where the spherical wave $S$ cuts through
the line connecting \(P\) and \(O\). We add another sphere---now
with a radius which is increased by half a wavelength---and continue this
procedure of slicing the spherical wavefront into many
layers. The bottom of Fig.~\ref{fig1} shows a two-dimensional cut through $S$.

The total strength at the point \(P\) is then equal to the alternating
sum \cite{r3}\begin{equation}
\label{psip}
\mathcal{U}(P)=\beta \sum ^{\infty }_{n=0}(-1)^{n}\mathcal{K}_{n}
\end{equation}
 where \(\beta \) is a factor and \emph{\(\mathcal{K}_{n}\)}
is the inclination factor evaluated at the angle \(\chi _{n}\)
appearing at the crossing of the \(n\)-th circle with the spherical
wave.

The similarity of the expressions for the Wigner function in terms of the energy distribution, Eq.~(\ref{w_alpha}), and the strength of a Fresnel wave in terms of inclination factors, Eq.~(\ref{psip}), is astonishing. But maybe it is nothing more than a formal mathematical coincidence.

Immediately three reasons why we are tempted to agree with this evaluation
of the situation offer themselves: (i) There does not seem to be an
analogue to the Fresnel zones in phase space. (ii) The Wigner function
lives in phase space whereas the Fresnel zones appear in real space.
(iii) There does not seem to be a general argument that explains
why Fresnel optics is preferred over other limits of optics.

In the next sections we address these objections one by one. In this process we will utilize one crucial tool, namely the semi-classical limit of quantum mechanics. It brings to light Fresnel zones in phase space, connects phase space with real space and, most importantly, singles out Fresnel optics from other limits of optics.

\section{A First Clue: Overlap in Phase Space \label{sec4}}

We start by showing that a picture reminiscent
of Fresnel zones emerges in phase space. Moreover, we give a geometrical meaning to the energy distribution \(P_{n}\)
in Eq.~(\ref{w_alpha}).

The concept of area of overlap in phase space \cite{r6,r7} is the bridge between the two apparently disconnected fields of Fresnel optics and phase space. Indeed, the rigorous approach is the Wigner function. It allows us to represent \cite{r10} the scalar product of two quantum states as the overlap in phase space.

However, a much more elementary algorithm springs from the picture
of an energy eigenstate as a band \cite{r6,r7} in Bohr-Sommerfeld
phase space. In this formalism the edges of the \(n\)-th band are
determined by the familiar Bohr-Sommerfeld quantization condition
\begin{equation}
\oint dx\, p(x)=2\pi \hbar n.
\end{equation}

\begin{figure}
\centering{\includegraphics{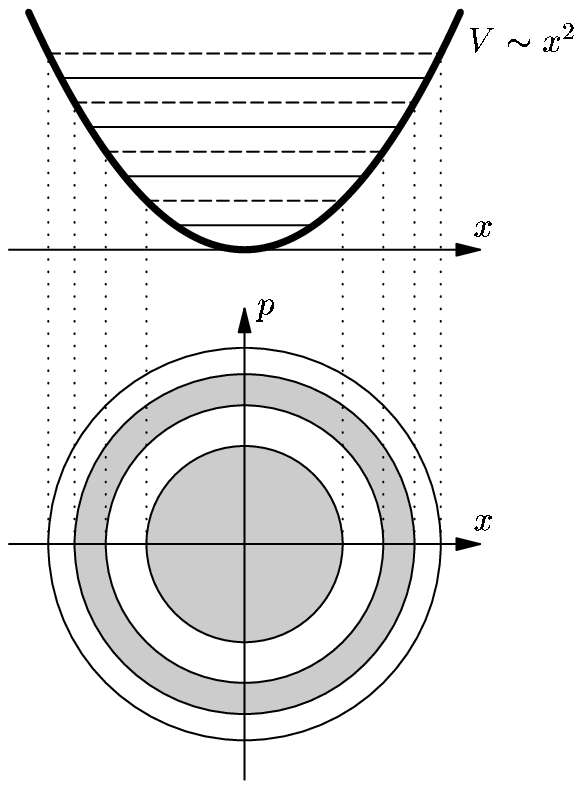}}
\caption{\label{fig2}\abovedisplayskip0.5ex\belowdisplayskip0.5ex Energy eigenstates of the harmonic oscillator represented as circular Bohr-Sommerfeld bands in phase space. Whereas the correct quantum mechanical energy levels depicted in the quadratic potential by solid lines are determined by the quantization condition $$\text{$\text{enclosed phase space area}=2\pi\hbar(n+1/2)$},$$ the edges of the Bohr-Sommerfeld bands are given by the energies following from the condition $$\text{$\text{enclosed phase space area}=2\pi\hbar n$}$$ as indicated by dashed lines.}
\end{figure}

For the case of a harmonic oscillator the edges of the \(n\)-th circular
Bohr-Sommer\-feld band are proportional to \(\sqrt{n}\) as shown in Fig.~\ref{fig2}. In this way we discover a picture
that is reminiscent of the Fresnel zones of Fig.~\ref{fig1}.

\begin{figure}
\centering{\includegraphics{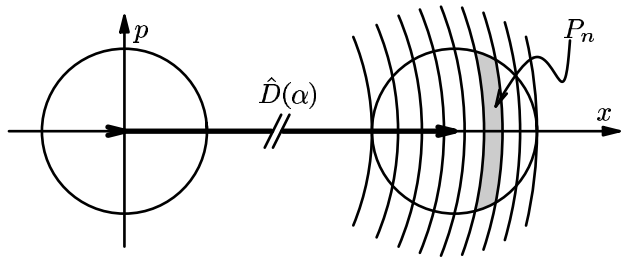}}
\caption{\label{fig_overlap}Energy distribution $P_n$ of a coherent state $\ket\alpha\equiv\hat D(\alpha)\ket0$ as area of overlap in phase space. The probability $P_n$ to find the $n$-th energy eigenstate in the coherent state is the area of overlap \cite{r1} between the coherent state represented by a displaced circle and the $n$-th Bohr-Sommerfeld band. This algorithm gives rise to a half-circle energy distribution rather than the exact Poissonian. A more sophisticated approach recognizes the Gaussian internal structure of the coherent state which leads to the Gaussian limit of the Poissonian.}
\end{figure}

In this formalism the energy distribution $P_n$ of an arbitrary state is the overlap between the state represented appropriately in phase space and the Bohr-Sommerfeld bands. For the example of a coherent state, that is, a displaced ground state, it is thus given \cite{r1} by the overlap of a circle representing the coherent state with these bands as shown in Fig.~\ref{fig_overlap}. It is interesting to note that \cite{r1} already mentions briefly a connection to Fresnel optics.

Hence, we can interpret the energy distribution
in the definition Eq.~(\ref{w_alpha}) of the Wigner function as the
overlap in phase space between circular bands and a displaced state in the
spirit of the Fresnel zone construction.

Still, there seems to be an immediate contradiction between this phase
space approach and the Fresnel picture. Indeed, the radii of the Fresnel
spheres increase linearly with \(n\) whereas the edges of the Bohr-Sommerfeld
bands increase only as the square root of \(n\).

Fortunately, here we focus on the wrong comparison. Instead of concentrating
on the radius of the sphere we should rather consider the radius of
the Fresnel zone. As indicated in the lower part of Fig.~\ref{fig1} this radius is the distance from
the crossing point between the Fresnel sphere and the wavefront $S$ to
the point on the line \(OP\). It increases with the square root of
\(n\). Thus the Fresnel zones enjoy the same scaling law as Bohr-Sommerfeld bands.

\section{Another Clue: From Real Space to Phase Space\label{sec5}}

\begin{figure}
\centering{\includegraphics{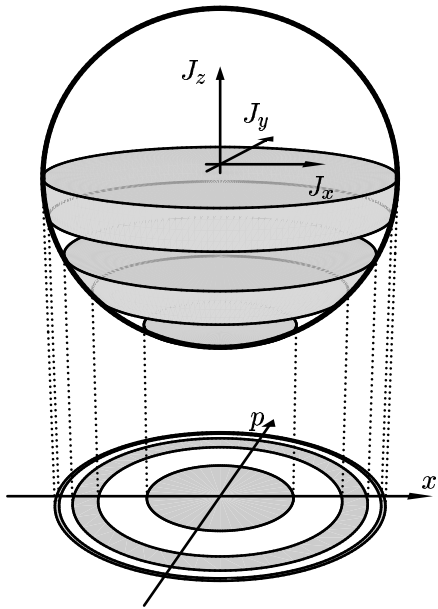}}
\caption{\label{fig3}From angular-momentum phase space to harmonic oscillator phase space via stereographic projection. In the semiclassical limit, that is, for $1\ll J$, the slices on the southern hemisphere representing states of well-defined $J_z$ values turn into the Bohr-Sommerfeld bands. This asymptotic connection breaks down for slices close the equator. Here the resulting Bohr-Sommerfeld bands pile up and their phase space areas are becoming smaller than $2\pi\hbar$. A similar picture holds for the states on the northern hemisphere.}
\end{figure}

Finally, there is still the lingering question of different spaces.
The Wigner function lives in phase space whereas Fresnel optics acts
in real space. To resolve this apparent contradiction we have to draw
on another analogy: The vector model of angular momentum.

We can represent the quantum state of an angular momentum of total
value \(J\) as a domain on a sphere of radius \(\sqrt{J(J+1)}\).
States of a well-defined \(z\)-component of \(\hat J\) play a prominent
role. In this case the domains on the sphere are \(2J+1\) belts as shown on the top of Fig.~\ref{fig3}. An angular-momentum vector with the \(z\)-component \(J_{z}=m\hbar \)
can be found anywhere in the \(m\)-th belt.

A more sophisticated approach defines the Wigner function of an angular-mo\-mentum state \cite{r8}. In complete analogy to the Wigner function
Eq.~(\ref{wigner_def}) in position and momentum we can again associate
positive and negative values with every point on the angular-momentum sphere.

We emphasize that this picture of an angular momentum is still in
real space. So where is phase space? Phase space for an angular momentum
is real space! Moreover, in the limit of large positive values of
\(J\) we can relate angular-momentum phase space to that of the
harmonic oscillator by making use of the stereographic projection
\cite{r15} shown in the bottom of Fig.~\ref{fig3}. In this way we project the Wigner function wrapped around
the sphere onto a plane. We find the familiar Wigner function
of an energy eigenstate of a harmonic oscillator. From phase space
of angular momentum living in real space we have gone to the phase
space of the harmonic oscillator.

\section{Conclusions\label{sec6}}

{}``Never do a calculation unless you know the answer.'' This advice
has always been and still is one of John Archibald Wheeler's first
moral principles in science. In the case of Wigner versus Fresnel
we have been good students of Wheeler. Indeed, we have an answer.
Moreover, we have many indications for the correctness of our hypothesis.
We have presented the evidence: Alternating sums define the Wigner
function as well as the strength of a Fresnel wave. The quantities
that enter the sums result from similar geometrical constructions.
They even scale similarly. One final and rather general argument in
favor of Fresnel optics comes to our mind. Again Wheeler is our witness.
As he points out \cite{r11} Fresnel optics and its quadratic phase
factors giving rise to the Cornu spiral play a crucial role in Feynman's
path integral formulation of quantum mechanics.

Despite the wealth of arguments the calculation is still missing and
may serve as an opportunity for a new collaboration with the birthday
boy. Happy birthday, Frank!

\newpage
\section*{Acknowledgments}

It is our great pleasure to thank R.~F.~O'Connell, J.~P.~Dahl, K.~Vogel, H.~Walther, and A.~Wolf for many fruitful discussions
on a multitude of interesting aspects of Wigner functions, Fresnel
optics and interference in phase space. In particular, W.~P.~S. is grateful
to J.~A.~Wheeler for introducing him to phase space. The generous support
of the Land of Baden-W\"{u}rttemberg is acknowledged.

\end{document}